\documentclass[onecolumn,pre]{revtex4}
\usepackage{epsfig}
\usepackage{graphicx}
\usepackage{epsfig}
\usepackage{dcolumn}
\usepackage{bm}
\begin{document}

\title{Statistical Mechanics of the Hyper Vertex Cover Problem}
\author{Marc M\'ezard and Marco Tarzia}
\affiliation{CNRS; Laboratoire de Physique Th\'eorique et Mod\`eles 
Statistiques, Univ. Paris-Sud, UMR 8626, 
Orsay, CEDEX, 91405 France}
\date{\today}

\begin{abstract}
  We introduce and study a new optimization problem called Hyper Vertex Cover.
  This problem is a generalization of the standard vertex cover to
  hypergraphs:  one seeks a configuration of particles with minimal density
  such that every hyperedge of the hypergraph contains at least one particle.
  It can also be used in important practical tasks, such as the Group Testing
  procedures where one wants to detect defective items in a large group by
  pool testing. Using a Statistical Mechanics approach based on the cavity
  method, we study the phase diagram of the HVC problem, in the case of random
  regualr hypergraphs. Depending on the values of the variables and tests
  degrees different situations can occur:  The HVC problem can be either in a
  replica symmetric phase, or in a one-step replica symmetry breaking one. In
  these two cases, we give explicit results on the minimal density of
  particles, and the structure of the phase space. These problems are thus in
  some sense simpler than the original vertex cover problem, where the need
  for a full replica symmetry breaking has prevented the derivation of exact
  results so far. Finally, we show that decimation procedures based on the
  belief propagation and the survey propagation algorithms provide very
  efficient strategies to solve large individual instances of the hyper vertex
  cover problem.
\end{abstract}

\maketitle

\section{Introduction and motivation}

The vertex cover problem is one of the standard NP-complete problems~\cite{NP}.
It is also intimately related to spin glass problems in statistical physics, 
and has received a lot of attention in recent years from the physics 
community~\cite{weigt,weigt_zhou,vc1,vc}.
In this paper we study a generalization of this problem to hyper-graphs, 
which we call the hyper vertex cover (HVC) problem. To the best of our 
knowledge this problem has not been studied before, at least not from a 
statistical physics perspective. The problem is easily stated:
We consider a large population of $N$ variables $x_i$, which can be either 
{\em active} ($x_i = 1$) or {\em inactive} ($x_i = 0$), and
$M$ function nodes (or {\em tests}), $t_a$, 
and build up a random regular hyper-graph (or {\em factor graph}),
i.e., a bipartite graph where each 
variable is connected exactly to $L$ tests, and each test 
is connected exactly to $K$ variables (with $NL = MK$).
The function nodes enforce the constraint that at least
one of the variables to which they are connected
must be active.
The case $K=2$ reduces to the standard
vertex cover (VC) problem.
In this paper we will be mainly interested in the {\em
minimal} HVC, i.e. a cover of the hyper-graph
with the minimal possible number of active variables.

The model can be also interpreted as a system of $M$ Boolean 
clauses over the $N$ variables which should be simultaneously
satisfied: 
each clause is an or function involving $K$
randomly chosen variables, (e.g., $x_{i_1} \lor \ldots
\lor x_{i_K} = 1$) and they are such that each variable
appears exactly in $L$ clauses. Thus, the minimal HVC
configuration corresponds to a pattern of the variables $x_i$ which
satisfies all the $M$ clauses with the minimal number of ones.

The interest of the present study is twofold. Statistical physics studies of
the VC problem on random graphs have shown that, depending on the average
degree of a variable in the graph, the problem is either simple (meaning that
replica symmetry is unbroken), or very difficult (meaning that a full replica
symmetry breaking - RSB - scheme is necessary). As the theory of full RSB is
not well under control for 'finite connectivity' graphs (where the degrees of
variables are finite), the difficult region is not fully understood,
notwithstanding the recent progress made in
Refs.~\cite{weigt,weigt_zhou,vc1}.
As we shall see the HVC displays, for certain families of random graphs, an
intermediate situation which is both non-trivial, because RSB is needed, but
under control, because the solution is given by a first order RSB (1RSB), as
developed for finite connectivity problems in~\cite{marc1}. Therefore this
model joins the family of well controlled hard combinatorial optimization
problems in which 1RSB is supposed to give exact results, a family which
includes $K$-satisfiability~\cite{marc2,SP,giulio,mon_zec}, 
graph coloring~\cite{coloring}, random Boolean
equations~\cite{XOR}, 1-in-$K$ satisfability~\cite{lenka}, and lattice 
glasses~\cite{lgm,rivoire}.

On the other hand, there are several applications of HVC
to important practical ``real-world'' tasks. In particular,
HVC is closely 
related to the {\em Group Testing} (or {\em Pool Testing}) 
procedures~\cite{GT,MT}.
The object of the Group Testing is to identify an a priori unknown
subset of a large population of $N$ variables, 
called the set of {\em active} (or {\em defective}) items, using
as few queries as possible. Each query (or test) is connected to a 
certain subset of $K$ items, and informs the tester about whether or not the
subset contains at least one active item. A negative answer implies that
all the items of the subset are inactive.
This approach is used in many different applications, beginning
with an efficient blood testing procedure~\cite{blood}. 
Other applications include quality control in product testing~\cite{quality},
searching files in a storage systems~\cite{files}, efficient accessing of 
computer memories, sequential screening of experimental
variables~\cite{screening}, and many others, such as the basic problem
of DNA libraries screening, which is very important in modern
biological applications such as mono-local antibody generation.
Objectives of group testing range from finding an optimal
strategy with the minimal number of tests, to devising an
efficient algorithm able to reconstruct the pattern of active items.

In  the Group Testing problem, it is easy to first identify
the variables which are  {\em sure zeros} (the
ones which are connected at least to one negative test). 
If one now considers the reduced graph obtained from the original GT problem by 
removing these sure zero variables, as well as all the negative tests,
one obtains a reduced graph (with fluctuating degrees). The problem of 
identifying the active items in this graph is exactly the HVC problem.
Therefore, the study of the phase diagram of the HVC
could give useful insights, for example, to understand
which is the best reconstruction algorithm for the pattern of active items
to be used depending on the topological properties of the factor graph.

In this paper we first present the study of the phase diagram of the random HVC 
problem, where each variable appears in $L$ tests and each test involves $K$
variables, based on the  cavity 
method~\cite{marc1} (these results could also be obtained
in the framework of the replica approach~\cite{SGT,giulio}).
We show that
depending on the values of the variables and tests degree
different situations can occur: The minimal HVC problem
can be either in a replica symmetric (RS) phase or in a one-step replica 
symmetry breaking (1RSB) phase. In these two cases
we give explicit results on the minimal density of active items, and the 
structure of the phase space. On the other hand, there are also cases (like, 
e.g., the ordinary VC with $K=2$ and $L\ge 3$), where a higher order RSB 
pattern is needed; these are more difficult problems that we don't address in 
this paper. The summary of the situation for the various  values of $K$ and $L$
is contained in Figure~\ref{fig:PD}. 
We then introduce the survey propagation (SP) and the belief propagation (BP) 
type algorithms, the analog for this problem of the ones which have turned out 
to be so efficient in $K$-satisfiability~\cite{SP}. We show that a decimation
procedure based on the surveys turns out to be an efficient way to solve large 
instances of the HVC problem.

The paper is organized as follows: In the next section we define the problem
and we introduce the statistical mechanics formulation;
In Sec.~\ref{sec:rs} we develop the cavity approach and work out the 
RS solution; In Sec.~\ref{sec:1rsb} we focus on the 1RSB solution for the phase diagram,
on the minimal HVC limit and on the stability of the 1RSB approach
with respect to further breaking of the Replica Symmetry;
Sec.~\ref{sec:SP} explains the use of the BP  and SP algorithms, and their 
application with a decimation procedure, to solve individual large instances of 
the problem; Finally, in Sec.~\ref{sec:concl} we discuss the results found
and conclude this work.

\section{Hyper vertex cover: definition and statistical physics formulation}
We consider a  factor graph containing $N+M$ vertices: $N$ of them are 
associated with variables, 
we shall label them by $i,j,...\in\{1,\dots,N\}$.
The other  $M$ are associated with function nodes (or tests), denoted by 
$a,b... \in\{  1 , \ldots , M\}$. An edge $i-a$ between variable $i$ and vertex 
$a$ is present in the factor graph  whenever  variable $i$ appears in test $a$. 
The factor graph is bipartite. 

The HVC problem is the following: each variable $i$ can be
either inactive ($x_i = 0$) or active ($x_i=1$). We request that, for each of
the $M$ tests, at least one out of the $K$ variables connected to it be equal
to one. The optimization problem (minimal HVC) consists in finding a 
configuration that satisfies all these constraints and has the 
smallest value $A_{\text{min}}$ of the ``weight'':
\begin{equation} \label{eq:A}
A(\{x_i\})=\sum_{i=1}^N x_i \ .
\end{equation}
 One is also interested in knowing the number
of configurations which satisfy all constraints, when $A$ has a fixed value 
$\ge A_{\text{min}}$.

We shall use the following statistical mechanics formulation of the problem.
Given an instance of the problem, characterized by a factor graph, we introduce
the set of admissible configurations, $ \mathcal{C}$, which are all
configurations of the $N$ variables such that, for each clause $a$, at least one
variable connected to $a$ takes value $1$. Denoting by $\partial a $ the
variables which enter in clause $a$ (i.e. the set of neighbors of $a$ in the
factor graph), we thus impose $\prod_{i \in \partial a} (1 - x_i) =0$.
The Boltzmann-Gibbs measure (canonical ensemble) is defined as
$P(\{x_i\})=(1/Z) e^{- \mu A (\{x_i\})}$. The chemical potential $\mu$
controls the overall density of ones $\rho = \sum_i x_i / N$. The minimal HVC
problem is recovered in the $\mu \to \infty$ limit, where the Boltzmann-Gibbs
measure concentrates on configurations with the smallest number of active
variables. We are also interested in understanding the properties of the
microcanonical measure $P^{mc}_A$ which is the uniform measure on the
admissible configurations $\{x_i\}$ with a fixed weight $A(\{x_i\})=A$,
whenever such configurations exist. These properties will be studied hereafter
through a detour to the canonical ensemble, using ensemble equivalence.

The partition function of the model reads:
\begin{equation}
Z = \sum_{\{x_i\}} e^{- {\cal H} (\{x_i\})} = e^{- \mu F} 
= e^{S - \mu N \rho},
\end{equation}
where $F$ is the free energy and $S$ is the entropy.

In the following we shall be particularly interested in the random $L,K$ HVC
problem, in which the factor graph is a random regular $(L,K)$ bipartite graph,
uniformly chosen from all the possible graphs where each variable is connected
to $L$ tests and each test is connected to $K$ variables.

The thermodynamic limit is taken by letting the number of variables $N$ and
the number of constraints $M$ go to infinity with a fixed ratio $\alpha=M/N$, 
keeping the degrees $K$ and $L$ fixed.
In the following we use the cavity method~\cite{marc1},
which allows to write down iterative self-consistent 
relations for local expectation values, which are exact on
a loop-less factorized graph (i.e., a tree).
For any finite values of $N$ and $M$ the graph is only locally
tree-like, and it has loops whose average length is expected to 
scale as $\ln N$. Therefore the cavity method is expected to provide
good approximations for large samples and to become exact
in the thermodynamic limit.

\section{Cavity approach and Replica Symmetric solution} \label{sec:rs}
\begin{figure}
\begin{center} \includegraphics[scale=0.85]{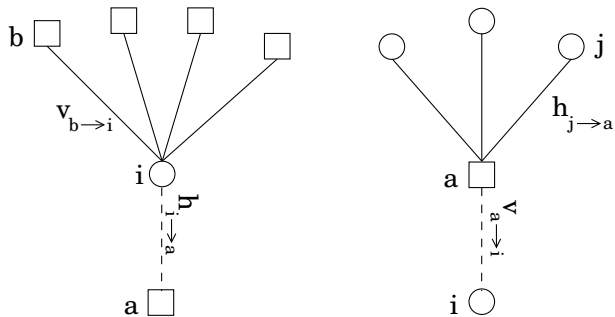}
\end{center}
\caption{Sub-graph rooted in $i$ in absence of the neighboring test $a$ (on the 
left). The function nodes $b$ belong to the neighborhood of $i$ ($b \in 
\partial i \setminus a$) and send the messages $v_{b \to i}$ to the site $i$.
The site $i$ sends the message $h_{i \to a}$ to the function node $a$, 
according to  
Eqs.~(\ref{eq:cavity}). Analogously, on the left we have sketched the sub-graph
rooted in the function node $a$ in absence of the edge with the variable $i$.
The sites $j$ belong to the neighborhood of $a$ ($j \in \partial a 
\setminus i$) and send the messages $h_{j \to a}$ to the function node $a$.
According to Eqs.~(\ref{eq:cavity}), $a$ sends the message $v_{a \to i}$
to the site i.}
\label{fig:graph}
\end{figure}

Given a graph and a variable $i$, we consider a sub-graph rooted in $i$
obtained by removing the edge between $i$ and one of its
neighboring tests, $a$.  Define $Z_0^{(i \to a)}$ 
and $Z_1^{(i \to a)}$ as the
partition functions of this sub-graph restricted to configurations where the
variable $x_i$ is respectively inactive ($x_i=0$) or active ($x_i=1$).
Assuming that the sub-graph is a tree (which is generically correct,
when one takes the large $N$ limit, up to any finite depth),
these restricted partition functions can be written
recursively (see Fig.~\ref{fig:graph}): consider the function nodes 
$b$ belonging to the neighborhood $\partial i$ of $i$.
For each $b\in \partial i\setminus a$, consider 
the restricted partition functions $Y^{(b \to i)}$ and $Y_1^{(b \to i)}$ on the rooted branches
of the graph starting from $b$, which are
defined, respectively, as the
total partition function of the branch, and the partition function
of the branch restricted to configurations in which at least one
of the remaining $K-1$ variables connected to the test is 
active.

One obtains the 
following recursion relations:
\begin{eqnarray} \label{eq:rs1}
\nonumber
Z_1^{(i \to a)} & = & e^{-\mu} \prod_{b \in \partial i \setminus a} 
Y^{(b \to i)}\\
Z_0^{(i \to a)} & = & \prod_{b \in \partial i \setminus a} 
Y_1^{(b \to i)} 
\end{eqnarray}
Analogously, one can express $Y^{(a \to i)}$ and $Y_1^{(a \to i)}$
in terms of the restricted partition functions $Z_0^{(j \to a)}$ 
and $Z_1^{(j \to a)}$  for $j\in\partial a \setminus i$:
\begin{eqnarray} \label{eq:rs2}
\nonumber
Y_1^{(a \to i)} & = & \prod_{j \in \partial a \setminus i} 
\left( Z_0^{(j \to a)} + 
Z_1^{(j \to a)} \right) - \prod_{j \in \partial a \setminus i} 
Z_0^{(j \to a)}\\
Y^{(a \to i)} & = & \prod_{j \in \partial a \setminus i} 
\left( Z_0^{(j \to a)} + 
Z_1^{(j \to a)} \right)
\end{eqnarray}
It is now convenient to introduce two local cavity fields on each edge of the 
graph, defined as:
$e^{\mu h_{i \to a}} = Z_1^{(i \to a)} / (Z_0^{(i \to a)} + Z_1^{(i \to a)})$, 
and $e^{\mu v_{a \to i}} = Y_1^{(a \to i)} / Y^{(a \to i)}$. Basically, 
$e^{\mu h_{i \to a}}$ measures the local probability that the variable $i$ is
active in absence of the link with the test $a$.
In terms of the local cavity fields, the recursion relations, 
Eqs.~(\ref{eq:rs1}) and (\ref{eq:rs2}), read:
\begin{eqnarray} \label{eq:cavity}
e^{\mu h_{i \to a}} &=& \frac{ \exp (- \mu ) } { \exp( - \mu ) + \exp \left( 
\sum_{b \in \partial i \setminus a} \mu v_{b \to i} \right) }\\
\nonumber
e^{\mu v_{a \to i}} & = & 1 - \prod_{j \in \partial a \setminus i}
\left( 1 - e^{\mu h_{j \to a}} \right).
\end{eqnarray}
For any given finite graph, 
Eqs.~(\ref{eq:cavity}) provide a set of $2MK$ (two equations for each edge 
of the graph) coupled algebraic equations, the so-called {\em belief 
propagation} (BP) equations, which can in principle be solved on any
individual instance.
From these local fields one can compute the system free energy
density $f = F/N$ in terms of {\em variable}, {\em test} and {\em edge}
contributions~\cite{marc1}. In order to do that, let us consider an intermediate
object, a factor graph made up by $N$ variables and $M$ tests
where $KL$ 'defective' variables are connected only
with $L-1$ tests and $KL$ 'defective' tests are connected only to $K-1$ 
variables, while all other variables and tests have their natural degree 
(respectively $L$ and $K$).
We can now go from this intermediate graph to a well defined regular 
factor graph
where each variable is connected to $L$ tests and each test to $K$ 
variables in two ways: 
\begin{itemize}
\item[i)] We can either add $K$ new items and $L$ new tests
and connect each of the item to $L$ out of the $LK$ defective tests and
each new test to $K$ out of the $LK$ defective variables. In this way
we obtain a regular hyper-graph made up by $N+K$ variables and
$M+L$ tests, all with their natural degree.
\item[ii)] Or we can add $LK$ new edges between pairs of defective variables
and tests. In this way we obtain a regular hyper-graph made up by $N$ variables
and $M$ function nodes, all with their natural degree.
\end{itemize}
In formulae we get:
\begin{equation} \label{eq:F1}
F(N+K) - F(N) = F_0 + \sum_{i=1}^K \Delta F_V^{i} 
+ \sum_{a=1}^L \Delta F_T^{a } - \left (
F_0 + \sum_{(j,b)} \Delta F_e^{(j,b)} \right),
\end{equation}
where $F_0$ is the free energy of the intermediate graph, 
$\Delta F_V^{i}$ is the free energy shift due to the
addition of a new variable $i$, $\Delta F_T^{a}$ is
the free energy shift due to the addition of a new test $a$, and
$\Delta F_e^{(j,b)}$ is the free energy shift due to the addition
of a new edge between the item $j$ and the test $b$.
Supposing that the free energy scales linearly with the number of 
items, the previous relation allows us to determine the free energy 
density.

The free energy shifts appearing in Eq.~(\ref{eq:F1}) 
can be written in terms of the restricted
partition functions defined above:
\begin{eqnarray}
\nonumber
e^{-\mu \Delta F_V^{i }} & = & 
\prod_{b \in \partial i} Y_1^{(b \to i)}
+ e^{- \mu} \prod_{b \in \partial i}
Y^{(b \to i)} \over \prod_{b \in \partial i}
Y^{(b \to i)}\\
e^{-\mu \Delta F_T^{a}} & = &
\prod_{j \in \partial a} \left( Z_0^{(j \to a)} + 
Z_1^{(j \to a)} \right)
- \prod_{j \in \partial a} Z_0^{(j \to a)}
\over \prod_{j \in \partial a} \left( 
Z_0^{(j \to a)} + Z_1^{(j \to a)} \right)\\
\nonumber
e^{-\mu \Delta F_e^{(j,b)}} & = &
Z_0^{(j \to b)} Y_1^{(b \to j)} + 
Z_1^{(j \to b)} Y^{(b \to j)} \over
Y^{(b \to j)} \left ( Z_0^{(j \to b)} + 
Z_1^{(j \to b)} \right).
\end{eqnarray}
The free energy shifts
can be finally rewritten in the following way 
in terms of the local cavity fields:
\begin{eqnarray} \label{eq:shift}
\nonumber
-\mu \Delta F_V^{i \cup \partial i} & = &
\ln \left[ \exp ( - \mu )  + \exp \left(
\sum_{b \in \partial i} \mu v_{b \to i}
\right) \right]\\
-\mu \Delta F_T^{a \cup \partial a} & = &
\ln \left[ 1 - \prod_{j \in \partial a}
\left ( 1 - e^{\mu h_{j \to a}} \right) \right]\\
\nonumber
-\mu \Delta F_e^{(j,b)} & = & \ln \left [ 
e^{\mu h_{j \to b}} + e^{\mu v_{b \to j}} - 
e^{\mu h_{j \to b} + \mu v_{b \to j}} \right].
\end{eqnarray}
The expectation value of the number of active variables on each site $i$
(also called {\em marginals} of the BP equations), can be obtained by
deriving the free energy with respect to the chemical potential 
leading to:
\begin{equation} \label{eq:dens}
\rho = \left \langle \frac{1}{N} \sum_i x_i \right \rangle 
= \frac{1}{N} \frac{\partial \mu F}
{\partial \mu} = \frac{1}{N} \sum_i \frac{\exp(-\mu)}
{\exp (-\mu) + \exp \left( \sum_{b \in \partial i} 
\mu v_{b \to i} \right)}
\end{equation}  

\subsection{Replica Symmetric solution, entropy crisis and stability}
Eqs.~(\ref{eq:cavity}) can be written for arbitrary graphs
and they provide exact marginal probability distributions
(and thus exact densities of active variables) only for 
loop-less trees.
They are particularly suited for very large random hyper-graphs, 
where, due to the local tree-like structure, they are expected to
provide exact results in the RS phase.

The simplest hypothesis one can make is the so
called {\em Replica Symmetric} (RS) Ansatz.
Assuming that there is a single state 
describing the equilibrium behavior of the system, one can look
for {\em factorized Replica Symmetric} solutions
of the cavity equations, where all the local fields
are equal on all the edges of the graph, i.e.,
$h_{i \to a} = h_{RS}$ and $v_{a \to i} = v_{RS}$, $\forall (i,a)$. 
Within this approximation, Eqs.~(\ref{eq:cavity}) reduce to:
\begin{equation} \label{eq:fRS}
\mu v_{RS} = \ln \left \{ 1 - \left[ \frac{ e^{\mu (L-1) v_{RS} } } 
{ e^{- \mu} 
+ e^{\mu (L-1) v_{RS} } } \right]^{K-1} \right \}.
\end{equation}
The free energy shifts reduce to node and edge independent quantities
and can be easily evaluated, along with all the thermodynamic
observables.

\begin{figure}
\begin{center}
\includegraphics[scale=0.35,angle=270]{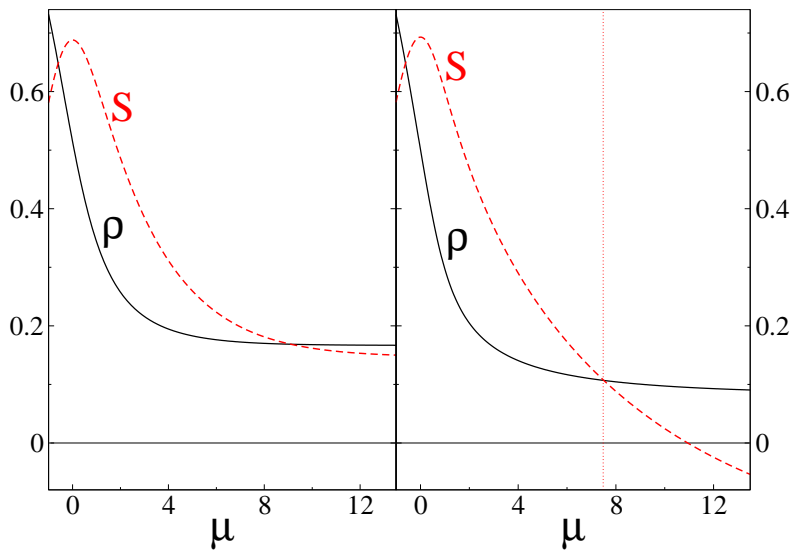}
\end{center}
\caption{Density of active items, $\rho$, and entropy density, $s = S/N$,
as a function of the chemical potential, $\mu$,
in the RS solution of the HVC for $L=2$ and $K=6$ (left panel)
and for $L=6$ and $K=12$ (right panel). The red
dotted vertical line in the right panel corresponds to the point 
where the RS solution becomes unstable}
\label{fig:rs}
\end{figure}

In Figure~\ref{fig:rs} the density of active variables $\rho$ and the entropy 
density $s = S/N = - \mu F/N + \mu \rho$ 
are plotted as  functions of the chemical potential $\mu$ for
$L=2$ and $K=6$ (left panel) and for $L=6$ and $K=12$ (right panel).
 In this RS solution, $\rho$  goes to $1/K$ as $\mu$
goes to infinity. 
One readily notes that while in the first case the entropy density
stays finite even in the $\mu \to \infty$ limit (meaning that there
is an extensive number of RS states with density of active
variables equal to $1/K$ satisfying the minimal covering of
the hyper-graph) in the second case the entropy density 
becomes negative for chemical potentials larger than a certain
threshold, $\mu_{s=0}$, (or, equivalently, for densities of active variables
smaller that a certain value).

These results can be easily understood by noting that in the $\mu \to \infty$ 
limit the RS fields are simply given by:
\begin{eqnarray}
\nonumber
\mu h_{RS} &=& - \, \frac{\mu}{L} - \frac{L-1}{L} \ln (K-1)\\
\mu v_{RS} & = & - \, \frac{\mu}{L} + \frac{1}{L} \ln (K-1).
\end{eqnarray}
Therefore, according to Eq.~(\ref{eq:dens}), $\rho = 1/K$, 
whereas the free energy density reads:
\begin{equation}
\mu f (\mu \to \infty) = 
\frac{\mu}{K} + \frac{(L-1)(K-1)-1}{K} \ln K - \frac{(L-1)(K-1)}{K} \ln 
(K-1),
\end{equation}
which immediately leads to:
\begin{equation}
s (\mu \to \infty) = \frac{1}{K}
\left \{ (L-1)(K-1) \ln (K-1) - \left[(L-1)(K-1) - 1 \right] \ln K \right \}.
\end{equation}
In the large connectivity limit ($K,L \gg 1$), the entropy density scales as:
\begin{equation}
s (\mu \to \infty; L,K \gg 1) \simeq \frac{\ln K - L}{K}.
\end{equation}
Thus, $s(\mu \to \infty)$ is positive 
if $\ln K > L$, while it is negative for $L > \ln K$.

Clearly, the results presented above
imply that the RS solution is wrong for $\mu$ large
enough, at least in the case $L=6$ and $K=12$. The RS solution turns out to be
incorrect if the assumption of the existence of a single equilibrium state
fails, meaning that fields incoming to a given node become correlated. To gain
further insight, one can test the stability of this RS solution by computing
the non-linear susceptibility, defined as:
\begin{equation}
\chi_2 = \frac{1}{N} \sum_{i,j} \langle x_i x_j \rangle_c^2.
\end{equation}
If $\chi_2$ diverges the incoming cavity fields are strongly correlated and the
RS assumption is inconsistent.
(Notice that the divergence of the linear susceptibility 
corresponds to a modulation instability which is not compatible with
the random nature of the graph).
By using the fluctuation-dissipation theorem, one can relate the connected
correlation functions between nodes $i$ and $j$ to the local cavity 
fields~\cite{rivoire}.
This finally leads to the following stability criterion:
\begin{equation}
\sum_{
j \in \partial a \setminus i;
b \in \partial j \setminus a}
\left( \frac{\partial v_{a \to i}}
{\partial v_{b \to j}} \right)^2 \leq 1.
\end{equation}
For the RS solution this criterion  yields:
\begin{equation}
\sqrt{(L-1)(K-1)} \left| \frac{e^{\mu h_{RS}} \left( 1 - e^{\mu v_{RS} } 
\right)} {e^{\mu v_{RS}} \left( 1 - e^{\mu h_{RS}} \right)} \right| 
\leq 1.
\end{equation}
Using the previous equation, we find that for several values of $L$ and $K$
the RS solution becomes indeed unstable above a certain chemical
potential. For $L=6$ and $K=12$ the point where the
instability appears is marked on the right plot as a vertical dotted line.
However, it is well known from the physics of glassy systems that the 
RS solution can be wrong because of a first order transition to a replica 
symmetry breaking solution, not detected by the stability argument.

\section{1RSB cavity approach} \label{sec:1rsb}
Figure~\ref{fig:rs} clearly shows that the RS solution fails at
high chemical potential and low density of
active items (at least for $L=6$ and $K=12$), due
to the fact that the hypothesis of the existence of a single
state becomes inconsistent.
Therefore, in this case other solutions must be 
found.
In the following we employ a one-step Replica Symmetry Breaking (1RSB)
approach~\cite{marc1} 
within the cavity method described in the previous
section.
More precisely, we assume that exponentially many ({\em pure})
states (in the
size of the system) exist and that the neighbors of a given node, 
in the absence of the node itself, are uncorrelated only within
each of these states.

Local fields on a given edge can now fluctuate from pure state 
to pure state.
The cavity methods provides a statistical
description of the local fields in each state $\alpha$, which must
be weighted accordingly to their Boltzmann weight 
$e^{- \mu F_{\alpha}}$~\cite{marc1}. 
In order to deal with this situation, 
on each edge of the graph
one has to introduce two probability distribution functions, 
$P_{i \to a} (h)$, and $Q_{a \to i} (v)$:
$P_{i \to a} (h)$ gives the probability density 
of finding the fields $h_{i \to a}$ equal to 
$h$ for a randomly chosen state
(respectively, $Q(v)$ gives the probability 
density of finding the field $v_{a \to i} = v$).
By using the cavity method, 
the following self-consistent integral 
relations are found~\cite{marc1,lgm}:
\begin{eqnarray} \label{eq:1RSB}
\nonumber
P_{i \to a} (h) &=& {\cal Z}_1 \int \prod_{b \in \partial i \setminus a} 
\left [ dv_{b \to i} Q(v_{b \to i})
\right] \, \delta \left( h - \overline{h} (\{v_{b \to i} \}) \right)
e^{- \mu m \Delta F_{V,iter}^{i \cup \partial i \setminus a} }\\
Q_{a \to i} (v) &=& {\cal Z}_2 \int \prod_{j \in \partial a \setminus i} 
\left [ dh_{j \to a} P(h_{j \to a})
\right] \, \delta \left( v - \overline{v} (\{h_{j \to a} \}) \right)
e^{- \mu m \Delta F_{T,iter}^{a \cup \partial a \setminus i} }
\end{eqnarray}
where $\overline{h} (\{v_{b \to i} \})$ and 
$\overline{v} (\{h_{j \to a} \})$
enforce the local cavity equations, Eqs.~(\ref{eq:cavity}), 
$m$ is the usual 1RSB parameter~\cite{SGT} (fixed by the maximization of
the free energy with respect to it), ${\cal Z}_1$ and ${\cal Z}_2$
are normalization factors, and 
$\Delta F_{V,iter}^{i \cup \partial i \setminus a}$ and 
$\Delta F_{T,iter}^{a \cup \partial a \setminus i}$ 
are the free energy shifts involved in the iteration processes,
which take into account the reweighting factors of the different
pure states.
Using the  relations:
\begin{eqnarray}
\nonumber
- \mu \Delta F_{V,iter}^{i \cup \partial i \setminus a} & = & \ln
\frac{ Z_0^{(i \to a)} + Z_1^{(i \to a)} }
{\prod_{b \in \partial i \setminus a}
Y^{(b \to i)} } = - \mu \Delta F_{V}^{i \cup \partial i}
+ \mu \Delta F_{e}^{(i,a)}\\
- \mu \Delta F_{T,iter}^{a \cup \partial a \setminus i}
& = & \ln \frac { Y^{a \to i} } {\prod_{j \in \partial a \setminus i} \left ( 
Z_0^{(j \to a)} + Z_1^{(j \to a)} \right) } = - \mu 
\Delta F_{T}^{a \cup \partial i} +
\mu \Delta F_{e}^{(i,a)},
\end{eqnarray}
one obtains  the expressions of the
free energy shifts involved in the iteration processes
in terms of the local cavity fields:
\begin{eqnarray}
- \mu \Delta F_{V,iter}^{i \cup \partial i \setminus a} &=& 
\ln \left[ \exp (-\mu ) + \exp \left( 
\sum_{b \in \partial i \setminus a} \mu v_{b \to i} \right) \right]\\
\nonumber
- \mu \Delta F_{T,iter}^{a \cup \partial a \setminus i} &=& 0.
\end{eqnarray}
In analogy with Eq.~(\ref{eq:F1}), 
from the local fields probability distribution, one can also compute
the 1RSB free energy of the system, as sum of contributions due to
the addition of a new item, $\Delta \phi_V^{i }$, 
addition of a new test, $\Delta \phi_T^{a}$, and
addition of a new edge between a variable and a test, $\Delta 
\phi_e^{(i,a)}$:
\begin{equation} \label{eq:FRSB}
K \phi = \sum_{i=1}^K \Delta \phi_V^{i } 
+ \sum_{a=1}^L \Delta \phi_T^{a } - 
\sum_{(j,b)} \Delta \phi_e^{(j,b)},
\end{equation}
with
\begin{eqnarray} \label{eq:1RSBfree}
\nonumber
\Delta \phi_V^{i } & = &
- \, \frac{1}{m \mu} \ln \left \{ \int 
\prod_{b \in \partial i} \left [ dv_{b \to i} Q(v_{b \to i})
\right] e^{- \mu m \Delta F_V^{i }} \right \}\\
\Delta \phi_T^{a } & = &
- \, \frac{1}{m \mu} \ln \left \{ \int
\prod_{j \in \partial a} \left [ dh_{j \to a} P(h_{j \to a})
\right] e^{- \mu m \Delta F_T^{a }} \right \}\\
\nonumber
\Delta \phi_e^{(i,a)} & = &
- \, \frac{1}{m \mu} \ln \left \{ \int
\left [ dh_{i \to a} P(h_{i \to a})
\right] \left [ dv_{a \to i} Q(v_{a \to i})
\right] e^{- \mu m \Delta F_e^{(i,a)}} \right \},
\end{eqnarray}
where the free energy shifts $\Delta F_V^{i \cup \partial i}$,
$\Delta F_T^{a \cup \partial a}$, and $\Delta F_e^{(i,a)}$ are 
defined in Eq.~(\ref{eq:shift}).

In the 1RSB formalism~\cite{marc1}, the 1RSB free energy, $\phi (\mu, m)$
is given by (in the thermodynamic limit):
\begin{equation}
- \mu m \phi (\mu, m) = - \mu m f(\mu) + \Sigma (f).
\end{equation}
Therefore, by Legendre transforming the 1RSB free energy, we obtain the
complexity $\Sigma (f)$ (i.e., the logarithm of the number of states with free
energy density equal to $f$):
\begin{equation} \label{eq:ltrsb}
\Sigma = \mu m^2 \frac{\partial \phi (\mu, m) }{\partial m}, \qquad
f = \frac{\partial [ m \phi (\mu, m) ] }{\partial m},
\end{equation}
with $\mu m = \partial \Sigma (f) / \partial f$.

On infinite random regular  hyper-graphs
one can assume that the 1RSB glassy phase is translationally
invariant, i.e., that the probability distribution of the local
cavity fields are edge independent: for all edges $a-i$,
$P_{i\to a}(h)=P(h)$ and $Q_{a \to i}(v)=Q(v)$.
In this case, Eqs.~(\ref{eq:1RSB}) become two coupled integral
equations for the probability distributions $P$ and $Q$, and can be solved for any
value of $\mu$ and $m$ by means of a 
population dynamics algorithm~\cite{marc1}. 

At low $\mu$ one finds with this procedure  that $P(h)$ and
$Q(v)$ always converge toward the RS solution: starting from 
populations of fields with an arbitrary distribution, they converge
to populations of identical fields such that
$P(h) = \delta (h - h_{RS})$ and
$Q(v) = \delta (v - v_{RS})$, where $h_{RS}$ and $v_{RS}$ are
the values of the local cavity fields which satisfy the factorized RS
equation, Eq.~(\ref{eq:fRS}).

When $\mu$ is increased, a first non-trivial distribution is found 
at $\mu = \mu_d$ for $m=1$. At this point many states appear.
The phase space splits into many clusters of solutions and,
even though they are only metastable (the equilibrium state
is still given by the RS solution at this point, since the maximum of the
1RSB free energy occurs at $m>1$), they could trap most of the
algorithms and dynamical procedures which look for a covering
pattern of a 
given hyper-graph. Thus, for a density of active items
smaller than this ``dynamical'' threshold a {\em survey propagation}
algorithm~\cite{marc2} 
should be used to find solutions of the HVC for finite instances.

A static phase transition (which is only relevant at equilibrium)
appears at higher chemical potential, $\mu = \mu_c$, where the 
maximum of the 1RSB free energy is located in $m=1$, the
complexity vanishes~\cite{marc1,lgm,rivoire}, and a thermodynamic transition from 
the RS phase to a 1RSB glassy one takes place (for $L=6$ and $K=12$
we find that $\mu_d \simeq 6.2$ and $\mu_c \simeq 7$). 

\subsection{Minimal Hyper Vertex Cover}
Now we consider the minimal HVC problem. Namely, we still
request that all the clauses are satisfied, but using the minimum possible number of
active variables.
This corresponds to the $\mu \to \infty$ limit
in our statistical physics formulation.

We thus consider  the $\mu \to \infty$ limit of the 1RSB equations, 
Eqs.~(\ref{eq:1RSB}).
In this limit, according to Eqs.~(\ref{eq:cavity}), it is
self-consistent to assume that the local cavity fields 
$h_{i \to a}$ and $v_{a \to i}$ 
can be either equal to minus one or to zero. 
In particular, the field $v_{a \to i}$ is equal to minus one if
all the incoming fields $h_{j \to a}$ are equal to minus one too, 
while it equals zero if at least one of the $h_{j \to a}$ is zero.
On the other hand, $h_{i \to a}$ turns out to be equal to minus one if
all the incoming fields $v_{b \to i}$ are equal to zero, and it equals
zero if at least one of the $v_{b \to i}$ is minus one.
Therefore, the probability distribution functions reduce to:
\begin{eqnarray}
\nonumber
P_{i \to a} (h) &=& g_{-1}^{i \to a} \delta (h + 1) + g_0^{i \to a} 
\delta (h)\\ 
Q_{a \to i} (v) & = & u_{-1}^{a \to i} \delta (v + 1) + u_0^{a \to i} \delta (v)
\end{eqnarray}
(with the constraints $u_0^{a \to i} + u_{-1}^{a \to i} = 1$ and 
$g_{0}^{i \to a} + g_{-1}^{i \to a} = 1$)
and the 1RSB integral equations reduce to algebraic equations for the
coefficients. For instance the second of Eqs.~(\ref{eq:1RSB}) becomes:
\begin{equation} \label{eq:sp1}
u_{0}^{a \to i} = \left [ 1 - \prod_{j \in \partial a 
\setminus i} \left ( 1 - g_{0}^{j \to a} \right) \right]
\end{equation}
where $y = \mu m$.
Analogously, for the first equation we have:
\begin{equation} \label{eq:sp2}
g_{0}^{i \to a} = \frac{ \left ( 1 - \prod_{b \in \partial i \setminus a}
u_{0}^{b \to i} \right ) e^{-y} }
{ e^{-y} + \left(1 - e^{-y} \right) \prod_{b \in \partial i \setminus a}
u_{0}^{b \to i} }.
\end{equation}
Finally, after some algebra, 
the following system of equations for the coefficients
$u_0^{a \to i}$ can be obtained: 
\begin{equation} \label{eq:1rsbu0}
u_0^{a \to i} = 1 - \prod_{j \in \partial a \setminus i} \left( \frac{\prod_{b \in \partial j \setminus a}
u_0^{b \to j}}{e^{-y} + \left( 1 - e^{-y} \right ) \prod_{b \in \partial j \setminus a} u_0^{b \to j}} \right).
\end{equation}
These equations are the (``zero temperature'') survey propagation (SP)
equations~\cite{SP,marc2} for the minimal HVC problem. 
In the case $K=2$, one recovers the result found for the VC in 
Refs.~\cite{weigt,weigt_zhou}.
We will see in the next section how to use them algorithmically in order to 
solve single instances.
Assuming that the coefficients $u_0^{a \to i}$ solving (\ref{eq:1rsbu0}) 
have been determined, the 1RSB free energy can be computed, 
according to Eqs.~(\ref{eq:1RSBfree}):
\begin{eqnarray}
\nonumber
\Delta \phi_V^{i } & = &
- \, \frac{1}{y} \ln \left[ 1 + \left( 1 - e^{-y} \right) 
\prod_{a \in \partial i} u_0^{a \to i} \right]\\
\Delta \phi_T^{a } & = & - \, \frac{1}{y} \ln \left[ 1 - 
\left( 1 - e^{-y} \right) \prod_{i \in \partial a}
\frac{\prod_{b \in \partial i \setminus a} u_0^{b \to i} }
{ e^{-y}  + \left( 1 - e^{-y} \right) \prod_{b \in \partial i \setminus a} 
u_0^{b \to i} } \right]\\
\nonumber
\Delta \phi_e^{(i,a)} & = & - \, \frac{1}{y} \ln \left[
1 - \left( 1 - e^{-y} \right)  \left( 1 - u_0^{a \to i} \right)
\frac{\prod_{b \in \partial i \setminus a} u_0^{b \to i} }
{ e^{-y}  + \left( 1 - e^{-y} \right) \prod_{b \in \partial i \setminus a} 
u_0^{b \to i} } \right]
\end{eqnarray}
In the minimal HVC limit one has that $- y \phi = \Sigma - y \rho$.
According to Eq.~\ref{eq:ltrsb}, the complexity $\Sigma$ 
is recovered by Legendre transforming the
function $\phi$ via the relation:
\begin{equation} \label{eq:mvc1}
\Sigma = y^2 \frac{\partial \phi}{\partial y},
\end{equation}
which leads to the following equation
for the density of active items:
\begin{equation} \label{eq:mvc2}
\rho = \phi + y \frac{\partial \phi}{\partial y}.
\end{equation}

\begin{figure}
\begin{center}
\includegraphics[scale=0.35,angle=270]{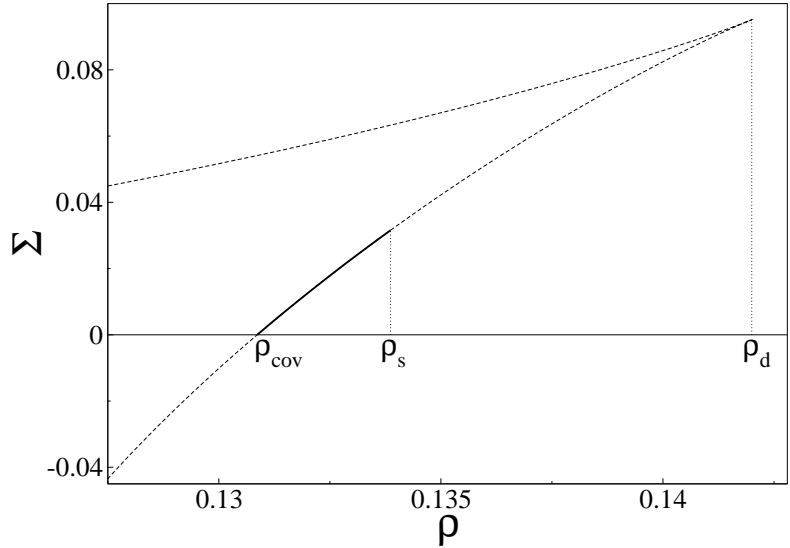}
\end{center}
\caption{Complexity $\Sigma$ as a function of the density
of active variables $\rho$ for $L=4$ and $K=8$. For
$\rho_{cov} \le \rho \le \rho_s$ the 1RSB Ansatz is stable. 
$\rho_{cov}$ (where $\Sigma = 0$) is the minimal covering density. Below 
$\rho_{cov}$ the complexity becomes negative: is no longer
possible to find solutions with smaller densities where all the clauses 
are satisfied and a COV/UNCOV transition takes place. 
The 1RSB Ansatz is no longer stable for $\rho>\rho_s$, where further breaking
of the replica symmetry are expected to occur.}
\label{fig:RSB}
\end{figure}

On infinite random-regular hyper-graphs, the coefficients$ u_0^{a \to i}$ can
be assumed to be edge-independent: $ u_0^{a \to i}=u_0$. In this case the
problem reduces to solving the single algebraic equation for the coefficient
$u_0$:
\begin{equation} \label{eq:1rsbu0_rr}
u_0 = 1 - \left( \frac{u_0^{L-1}}{e^{-y} + \left( 1 - e^{-y} \right )u_0^{L-1}} \right)^{K-1}
\end{equation}

Once one has found the solution $u_0$ to this equation,  
the 1RSB free energy simplifies to:
\begin{equation}
\phi(y) = - \, \frac{1}{y} \ln 
\frac{1 + \left(1 - e^{-y}\right) u_0^L}
{\left(1 - u_0 \right)^{\frac{L}{K}}}.
\end{equation}
The complexity $\Sigma(\rho)$, and the density
$\rho$ are then easily evaluated using Eqs.~(\ref{eq:mvc1}) and 
(\ref{eq:mvc2}).

In Figure~\ref{fig:RSB} the complexity $\Sigma$ is plotted as a function of 
$\rho$ for $L=4$ and $K=8$. 
The complexity vanishes at $\rho_{cov}$, corresponding to
the minimal HVC density. At lower densities $\Sigma$ is negative, 
implying that for $\rho<\rho_{cov}$ no solutions satisfying all the clauses can
be found and a COV/UNCOV transition takes place. 
The complexity has a maximum at $\rho_d$ (given by $\partial^2 [y \phi(y)] 
= 0$). 
The curve also displays a non concave part for $y<y_d$, the physical 
interpretation of which has not yet been understood. 

\subsection{Stability of the 1RSB solution}
To determine whether the equilibrium state is really described by a 1RSB 
solution
or whether further replica symmetry breakings occur, one has to study the 
stability 
of the 1RSB solution. The stability analysis of the RS Ansatz investigates
if the RS state tends to split into exponentially many states. 
Since in the 1RSB phase the Gibbs measure is decomposed in a
cluster of different thermodynamic pure states~\cite{SGT}, there are two 
different kinds
of instabilities that might show up~\cite{montanari,rivoire}: 
i) either the states can aggregate into
different clusters (in order to study this first instability
one has to compute the inter-state susceptibility); 
ii) or each state can fragment in different
states (in order to study this second instability the intra-state 
susceptibility must be computed).

In the minimal HVC limit, the instability of the first kind can be easily
studied by computing the eigenvalue of the 
$(1 \times 1)$ Jacobian matrix associated with 
Eq.~(\ref{eq:1rsbu0})~\cite{montanari,rivoire}. 
Since the linear susceptibility is related to a 
modulation instability incompatible with the underlying structure of the
lattice, the non-linear (spin glass) susceptibility must be considered. 
Therefore, the criterion for the stability of the 1RSB Ansatz simply reads:
\begin{equation} \label{eq:s1rsb1}
\sum_{j \in \partial a \setminus i ; b \in \partial j \setminus a} 
\left( \frac {\partial u_0^{a \to i} } {\partial u_0^{b \to j} } \right) ^2 
\le 1,
\end{equation}
which yields:
\begin{equation}
\sqrt{(L-1)(K-1)} \left | \frac{ \left (1 - u_0 \right) e^{-y} }
{ u_0 \left[ e^{-y} + \left( 1 - e^{-y} \right) u_0^{L-1} \right] } \right| 
\le 1.
\end{equation}

To study the instability of second kind, instead, we consider a two-step
RSB like Ansatz of the form ${\cal Q} [Q] \simeq \sum_l u_l \hat{\delta}
[ Q(u) - \delta (u - l) ]$, and 
${\cal P} [P] \simeq \sum_l g_l \hat{\delta}
[ P(h) - \delta (h - l) ]$,
where the 1RSB states coincide with the 2RSB clusters
and the 2RSB states reduce to single configurations.
We want to compute the widening of $\delta Q = \sum_{m \neq l} \epsilon_m
(\delta (u - m) - \delta (u - l))$, which can be
written in terms of the widening of 
$\delta P = \sum_{q \neq r} \gamma_q (\delta (h - q) - 
\delta (h - r))$, and check whether or not it grows under
iteration. In order to do that we have to 
sum over all the perturbations of the cavity
fields on the neighboring sites which change the configuration of a given site
from $l$ to $m$, according to their Boltzmann weight
(for a general explanation see~\cite{montanari,rivoire}):
\begin{eqnarray}
\nonumber
u_l^{a \to i} \langle \epsilon_m \rangle_l
&=& \frac{1}{{\cal Z}_1} \sum_{\{ j_1, \ldots , j_{K-1} \} \in \partial a \setminus i ; 
(g^{j_1 \to a}, \ldots, g^{j_{K-1} \to a}) \to l }
g^{j_1 \to a} \cdots g^{j_{K-1} \to a} e^{(y_2 - y_1) \Delta 
\phi_{T,iter}^{a \cup \partial a \setminus i} (g^{j_1 \to a}, \ldots, g^{j_{K-1} \to a})}\\
&& \times
\sum_{w, q \neq g^{j_w \to a}, (g^{j_1 \to a}, \ldots,q,\ldots
g^{j_{K-1} \to a}) \to m} e^{y_2 \Delta 
\phi_{T,iter}^{a \cup \partial a \setminus i} (g^{j_1 \to a}, \ldots, q, \ldots, g^{j_{K-1} \to a})}
\langle \gamma_{g^{j_w \to a}} \rangle_q.
\end{eqnarray}
A similar equation which gives the $\langle \gamma_q \rangle_r$
in terms of the $\langle \epsilon_m \rangle_l$ can be derived.
The problem can then be rewritten solely in terms of the coefficients
$u_0$ and $u_1$. Using a transfer matrix representation we find:
$\langle \epsilon_0 \rangle_{-1} = (L-1)(K-1) \sum_{b,c=0,-1;b \neq c}
T_{(0,-1)(b,c)} \langle \epsilon_b \rangle_c$.
Therefore 1RSB solution is stable provided that 
the eigenvalue of $T$ of largest modulus, $\lambda$, is such that 
$(L-1)(K-1)\lambda \le 1$.
The transfer matrix $T$ reads (without losing of generality we can set $y_1=y_2=y$):
\begin{equation}
T = \frac{1}{{\cal Z}} \left (
\begin{array}{ccc}
0 & ~ & u_0^{(L-1)(K-1)}e^{-y}\\
u_0^{LK-L-K} (1 - u_0) & ~ & 0
\end{array}
\right),
\end{equation}
where the normalization is given by 
${\cal Z} = [e^{-y} + (1 + e^{-y}) u_0^{L-1}]^{K-1}$.
Therefore, the stability criterion of the 1RSB solution is given by:
\begin{equation} \label{eq:s1rsb2}
(L-1)(K-1) \frac { \sqrt{
e^{-y} u_0^{2(L-1)(K-1)-1}(1-u_0) } }
{ \left[ e^{-y} + \left(1 - e^{-y} \right) u_0^{L-1} \right]^{K-1} } \le 1.
\end{equation}
For some values of $L$ and $K$, according to the stability criterions given
in Eqs.~(\ref{eq:s1rsb1}) and (\ref{eq:s1rsb2}), we find that the 1RSB solution
is stable around the COV/UNCOV transition, and therefore the threshold is
likely to be exact. On the contrary, for other values of $L$ and $K$ the
1RSB approach is unstable, and a higher order Replica Symmetry Breaking transition 
is expected to occur. In this case a more involved analysis would be 
required to locate the COV/UNCOV threshold (the 1RSB result is expected to provide
a lower bound). This happens, for instance, for $K=2$, where the
results already found for the standard VC problem are fully 
recovered~\cite{weigt}.

\begin{figure}
\begin{center}
\includegraphics[scale=0.35,angle=270]{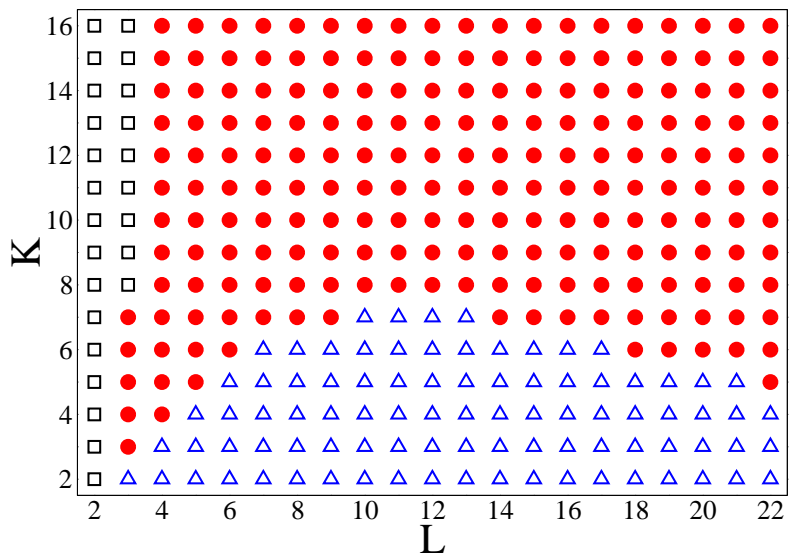}
\end{center}
\caption{Phase diagram of the Minimal Hyper Vertex Covering problem.
Black squares, red circles and blue triangles correspond, respectively,
to the values of $L$ and $K$ for which the minimal HVC configurations
are described by a Replica Symmetric Ansatz, a 1-step Replica Symmetry 
Breaking Ansatz, and a higher order Replica Symmetry Breaking Ansatz. 
As already shown in~\cite{weigt}, for $K=2$ the solution of
VC exhibit higher order RSB for every value of $L$.}
\label{fig:PD}
\end{figure}

The phase diagram of the system as a function of the item and test degrees
$L$ and $K$ is presented in
Figure~\ref{fig:PD}, showing the relative positions of the different
phases in the minimal covering limit.

\section{Survey propagation and survey inspired decimation}
\label{sec:SP}
We have already mentioned the possibility that, in analogy with 
$K$-satisfability and other optimization problems, the BP and SP equations, 
Eqs.~(\ref{eq:cavity}) and (\ref{eq:sp1})-(\ref{eq:sp2}), may
provide efficient algorithmic tools
to find solutions of the HVC problem on large
instances~\cite{SP,marc2}.
In the present section, taking advantage of the analytical investigations 
presented above, we test numerically the efficiency 
of these message passing algorithms on large samples (for several
values of the connectivity), and compare their performance to a
simple heuristic covering algorithm. We show that BP and SP provide an efficient
way to solve the HVC, being able to find covering
patterns whose sizes are very close to the minimal one. Moreover, in the
region of $K$ and $L$ where a 1-step RSB occurs, we show that an SP algorithm
allows to improve the BP results, and to find solutions of the HVC 
extremely close to the COV/UNCOV threshold predicted by the theoretical 
analysis.

In the following, we briefly describe the three algorithms used, and finally
we present the results found.

\subsection{Belief Propagation (BP) Algorithm}

The BP algorithm simply consists in finding by iteration a solution of 
Eqs.~(\ref{eq:cavity}) 
on a given factor graph, and then applying an iterative 
{\em decimation} procedure which, at each step, 
polarizes the most biased variables, until all the
variables are fixed. Thus, the BP algorithm works by iterating the three
following steps:
\begin{itemize}
\item[1.] Solve the BP equations, Eqs.~(\ref{eq:cavity}), on the graph, 
until all the messages converge to a fixed point.
\item[2.] Compute the marginals acting on each variable (i.e., the 
probability of being zero or one), and we polarize the most biased variable, 
by assigning to it the most probable value.
\item[3.] Generate the new {\em reduced graph}, by removing the variable
that has been polarized and all the edges incoming on it.
In the case where the variable has been polarized to one, we also remove 
all the tests to which it is connected (since they are automatically 
satisfied) and all the edges incoming on those tests.
\end{itemize}
There is a problem concerning the convergence of BP that has to be mentioned.
In particular, as discussed in the previous sections, depending on the value of 
$K$, $L$ and $\mu$, the entropy of the RS solution of the HVC (i.e., the number
of solutions of the BP equations) can be either positive or negative. 
In order to make the BP equations converge, we fix the value of $\mu$
in such a way that RS entropy of the problem is positive.
As a matter of fact, 
while the BP algorithm is iterated, the decimated graph modifies
and consequently the entropy associated to the problem defined on the reduced
graph changes. As a consequence, it can occur that the BP equations do
not converge on the reduced graph, after that some variables have been fixed.
In order to overcome this problem, during the decimation procedure 
we tune the chemical potential
in such a way that the RS entropy is always kept positive. (Another possible
and equivalent strategy consists in choosing at each decimation step the 
largest value of $\mu$ for which the BP equations converge to a fixed point).
\subsection{Survey Propagation (SP) Algorithm}
In some regions of the phase space the BP equations possess a high
number of solutions (corresponding to different thermodynamic states), and none 
can be found using a local iterative updating scheme.
BP works well if a single cluster of minimal HVC exists.
However, a breaking of the replica symmetry implies the
emergence of clustering in the solution space. This effect is captured
by the SP algorithm, as first proposed in~\cite{SP}, which describes
the statistics over all the solutions of the BP equation, by taking into
account their thermodynamic weight.
Basically, the SP algorithm is very much alike the BP one, and it consists
of the same three steps as before. The only difference
is that the SP messages, Eqs.~(\ref{eq:sp1}) and (\ref{eq:sp2}), 
must now be used instead of the BP ones.
In order to have a minimal HVC covering of the factor graph, 
we would like to set 
the value of $y$ to that corresponding to the COV/UNCOV transition, where
the 1-RSB free energy has a maximum and the complexity vanishes.
However, it can occur that, once that some variables have been fixed, 
the SP equations stop to converge on the reduced graph for that value of
$y$. This is due to the fact that while the decimation is carried on, the graph
and, consequently, the complexity change, and the value of $y$ one was
using may now fall into the uncoverable region of the decimated problem. 
Therefore, in order to overcome this problem, after some decimation
steps we recompute the complexity of the decimated problem defined on 
the reduced graph, and we tune
the value of $y$ to that corresponding of the new COV/UNCOV threshold.

\subsection{Greedy Algorithm}
Here we present a very simple heuristic algorithm which allows to find a 
covering pattern of the hyper-graph. The algorithm
consists of the following steps:
\begin{itemize}
\item[1.] We pick up the variable with higher degree and we set it to one.
We then remove
from the graph this variable and all the edges incoming on it. We
also remove all the tests 
connected to that variable (since they are satisfied) and all the edges
incoming on those tests.
We repeat this procedure until there are no variables left with degree 
larger than zero, and no more tests.
\item[2.] We fix to zero all the remaining variables, which are all isolated. 
\end{itemize}

\subsection{Results}
\begin{figure}
\begin{center}
\includegraphics[scale=1.]{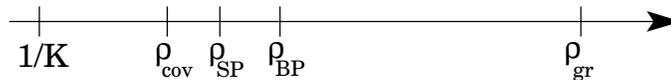}
\end{center}
\caption{Sizes of the covering of the hyper-graph obtained by
the BP, the SP and the greedy algorithm for a HVC problem with
$L=4$ and $K=6$, for $12288$ variables 
and $8192$ tests. The COV/UNCOV threshold in this case is equal to
$\rho_{cov} \simeq 0.178$; The SP algorithm allows to find
solutions of size $\rho_{SP} \simeq 0.182$; The BP algorithm
generates coverings of size $\rho_{BP} \simeq 0.186$. Finally
the greedy algorithm gives solutions of size $\rho_{gr}
\simeq 0.212$.}
\label{fig:algo}
\end{figure}

We have tested these three algorithms on large instances for several values 
of $L$ and $K$. It turns out that in general 
both BP ans SP perform much better than
the greedy algorithm, and are able to find efficiently
solutions of the HVC very close to the minimal COV/UNCOV threshold predicted
by the analytical calculations (in general the size of the 
solutions provided by BP and SP is always less than few per cents 
larger than the
minimal one). In particular, for values of $L$ and $K$ for which a 1-step RSB
transition occurs (see Fig.~\ref{fig:PD}), SP is able to improve the BP result.
Just to give a more quantitative idea of the performance of the different 
algorithms, in Fig.~\ref{fig:algo}, 
we present the results of the numerical tests of the three algorithms 
for $L=4$ and $K=6$.

\section{Discussion and Conclusion} \label{sec:concl}
In this paper we have studied the statistical mechanics of a generalized
Vertex Covering problem on the hyper-graph, which might have many
practical applications. As an example, the problem of Group Testing
is deeply related to the HVC and, the knowledge of the phase
diagram of the latter, could give important insights on how to devise an
efficient reconstruction algorithm for the former, depending on $L$, $K$ and
on the density of active items.

The minimal HVC has been studied in great detail. 
For low enough degree of the items ($L=2$, $L=3$ and $K>7$, $L=4$ and $K>21$, 
\ldots) we find that the RS solution describes correctly the minimal
covering configurations. For bigger values of $L$, the RS Ansatz becomes 
inconsistent, whereas 
the 1RSB solution is stable and is likely to provide the correct 
solution around the COV/UNCOV threshold. Both in the RS and in the 1RSB region 
we have found explicit results on the minimal density of active items required 
to cover the factor graph, and on the structure of the phase space.
On the other hand, there are also cases (e.g., the ordinary VC 
problem~\cite{weigt,weigt_zhou}, 
$K=2$ and $L \ge 3$) where further 
RSB steps are required.
In these cases 1RSB approach becomes inconsistent and higher RSB patterns 
are required.


Finally, we have shown that a decimation procedure based on the BP and SP
equations turns out to be a very efficient strategy to solve large
individual instances of the HVC problem.

\acknowledgments

We warmly thank L. Zdeborov\'a for kind and fruitful discussions.
The numerical computations were done on the cluster EVERGROW (EU consortium
FP6 IST). This work was partially supported by EVERGROW (EU consortium FP6 
IST).


\begin{thebibliography}{40}
\bibitem{NP} S. Cook, {\em Proc. of the third annual ACM symposium on Theory of 
computing}, 151-158, 1971; M. R. Garey and D. S. Johnson, {\em Computers and 
Intractability:A Guide to the Theory of NP-Completeness}, WH Freeman and Co. 
New York, NY, USA, 1979.

\bibitem{weigt} M. Weigt and A. K. Hartmann, Phys. Rev. E {\bf 63}, 056127 
(2001); M. Weigt and A. K. Hartmann, Phys. Rev. Lett. {\bf 84}, 6118 (2000).

\bibitem{weigt_zhou} M. Weigt and H. Zhou, Phys. Rev. E {\bf 74}, 046110 (2006).

\bibitem{vc1} M. Bauer and O. Golinelli, Eur. Phys. J. B {\bf 24}, 339 (2001);
H. Zhou, Eur. Phys. J. B {\bf 32}, 265 (2003);
H. Zhou, Phys. Rev. Lett. {\bf 94}, 217203 (2005).

\bibitem{vc} A. Frieze, Discr. Math. {\bf 81}, 171 (1990);
P. Gazmuri, Networks {\bf 14}, 367 (1984).

\bibitem{marc1}
M. M\'ezard and G. Parisi, Eur. Phys. J. B {\bf 20},  217 (2001);
M. M\'ezard and G. Parisi, J. Stat. Phys. {\bf 111}, 1 (2003).

\bibitem{marc2}
M. M\'ezard, G. Parisi, and R. Zecchina, Science {\bf 297}, 812 (2002).

\bibitem{mon_zec} R. Monasson and R. Zecchina, Phys. Rev. Lett. {\bf 76} (1996) 3881.

\bibitem{giulio}
G. Biroli, R. Monasson, and M. Weigt, Eur. Phys. J. B {\bf 14}, 551 (2000).

\bibitem{SP}
M. M\'ezard and R. Zecchina, Phys. Rev. E66 (2002) 056126.

\bibitem{coloring} F. Krzakala, A. Pagnani, and M. Weigt, 
Phys. Rev. E {\bf 70}, 046705 (2004);
R. Mulet, A. Pagnani, M. Weigt, and R. Zecchina, Phys. Rev. Lett. {\bf 89} 
268701 (2002); L. Zdeborov\'a and F. Krzakala, cond-mat/0704.1269.

\bibitem{XOR} M. M\'ezard, F. Ricci-Tersenghi, and R. Zecchina, 
J. Stat. Phys. {\bf 111}, 505 (2003).

\bibitem{lenka} J. Raymond, A. Sportiello, and L. Zdeborov\'a, 
cond-mat/0702610.

\bibitem{lgm}
G. Biroli and M. M\'ezard, Phys. Rev. Lett. {\bf 88}, 025501 (2002);
A. Hartmann and M. Weigt, Europhys. Lett. {\bf 62}, 533 (2003);
M. P. Ciamarra, M. Tarzia, A. de Candia, and A. Coniglio, Phys. Rev. E 
{\bf 67}, 057105 (2003).

\bibitem{rivoire}
O. Rivoire, G. Biroli, O. Martin, and M. M\'ezard, Eur. Phys. J. B {\bf 37}, 
55 (2004).

\bibitem{GT} T. Berger, IEEE Trans. Inform. Theory {\bf 48}, 1741 (2002); 
H. Q. Ngo and D.-Z. Du, DIMACS Series in Discrete Mathematics and
Theoretical Computer Science, (2000).

\bibitem{MT}
M. M\'ezard and C. Toninelli, arXiv:0706.3104.

\bibitem{blood}
D. Dorfman, Ann. Math. Statist. {\bf 14}, 436 (1943).

\bibitem{quality}
M. Sobel and P. A. Groll, Bell Syst, Tech. J. {\bf 38}, 1179 (1959).

\bibitem{files}
W. H. Kautz and R. R. Singleton, IEEE Trans. Inform . Theory {\bf 10}, 363 
(1964).

\bibitem{screening}
C. H. Li, J. Amer. Statist. Assoc. {\bf 57}, 455 (1962).

\bibitem{SGT}
M. M\'ezard, G. Parisi, and M. A. Virasoro, {\em Spin-glass Theory and Beyond}, 
vol. 9 of Lecture notes in Physics, World Scientific, Singapore, 1987.

\bibitem{montanari}
A. Montanari and F. Ricci-Tersenghi, Eur. Phys. J. B 33, 339 (2003)

\end{thebibliography}
\end{document}